\documentclass[preprint]{aastex631}
\usepackage{multirow}

\newcommand\exoplanet{\textsf{exoplanet}}

\newcommand\lsun{L$_\sun$}
\newcommand\mearth{M$_\earth$}
\newcommand\msun{M$_\sun$}

\newcommand\prot{P_{\rm rot}}

\newcommand\pymc{\textsf{PyMC3}}
\newcommand\rearth{R$_\earth$}
\newcommand\rsun{R$_\sun$}
\newcommand\searth{S$_\earth$}
\newcommand\teff{T_{\rm eff}}
\newcommand\teq{T_{\rm eq}}
\newcommand\tfull{T_{\rm full}}

\received{\today}
\submitjournal{AJ}

\begin{document}

\title{LHS 475 b: A Venus-sized Planet Orbiting a Nearby M Dwarf}

\correspondingauthor{Kristo Ment}
\email{kristo.ment@cfa.harvard.edu}

\author[0000-0001-5847-9147]{Kristo Ment}
\affiliation{Center for Astrophysics \textbar~Harvard \& Smithsonian, 60 Garden Street, Cambridge, MA 02138, USA}

\author[0000-0002-9003-484X]{David Charbonneau}
\affiliation{Center for Astrophysics \textbar~Harvard \& Smithsonian, 60 Garden Street, Cambridge, MA 02138, USA}

\author{Jonathan Irwin}
\affiliation{Center for Astrophysics \textbar~Harvard \& Smithsonian, 60 Garden Street, Cambridge, MA 02138, USA}

\author[0000-0001-6031-9513]{Jennifer G. Winters}
\affiliation{Center for Astrophysics \textbar~Harvard \& Smithsonian, 60 Garden Street, Cambridge, MA 02138, USA}
\affiliation{Williams College, 880 Main Street, Williamstown, MA 01267, USA}

\author[0000-0002-1533-9029]{Emily Pass}
\affiliation{Center for Astrophysics \textbar~Harvard \& Smithsonian, 60 Garden Street, Cambridge, MA 02138, USA}

\author[0000-0002-1836-3120]{Avi Shporer}
\affiliation{Department of Physics and Kavli Institute for Astrophysics and Space Research, Massachusetts Institute of Technology, Cambridge, MA 02139, USA}

\author[0000-0002-2482-0180]{Zahra Essack}
\affiliation{Department of Physics and Kavli Institute for Astrophysics and Space Research, Massachusetts Institute of Technology, Cambridge, MA 02139, USA}
\affiliation{Department of Earth, Atmospheric and Planetary Sciences, Massachusetts Institute of Technology, Cambridge, MA 02139, USA}

\author[0000-0001-9786-1031]{Veselin B. Kostov}
\affiliation{NASA Goddard Space Flight Center, 8800 Greenbelt Road, Greenbelt, MD 20771, USA}
\affiliation{SETI Institute, 189 Bernardo Ave, Suite 200, Mountain View, CA 94043, USA}

\author[0000-0001-9269-8060]{Michelle Kunimoto}
\affiliation{Department of Physics and Kavli Institute for Astrophysics and Space Research, Massachusetts Institute of Technology, Cambridge, MA 02139, USA}

\author[0000-0001-8172-0453]{Alan Levine}
\affiliation{Department of Physics and Kavli Institute for Astrophysics and Space Research, Massachusetts Institute of Technology, Cambridge, MA 02139, USA}

\author[0000-0002-6892-6948]{Sara Seager}
\affiliation{Department of Physics and Kavli Institute for Astrophysics and Space Research, Massachusetts Institute of Technology, Cambridge, MA 02139, USA}
\affiliation{Department of Earth, Atmospheric, and Planetary Sciences, Massachusetts Institute of Technology, Cambridge, MA 02139, USA}
\affiliation{Department of Aeronautics and Astronautics, Massachusetts Institute of Technology, Cambridge, MA 02139, USA}

\author[0000-0001-6763-6562]{Roland Vanderspek}
\affiliation{Department of Physics and Kavli Institute for Astrophysics and Space Research, Massachusetts Institute of Technology, Cambridge, MA 02139, USA}

\author[0000-0002-4265-047X]{Joshua N. Winn}
\affiliation{Department of Astrophysical Sciences, Princeton University, Princeton, NJ 08544, USA}

\begin{abstract}

Based on photometric observations by TESS, we present the discovery of a Venus-sized planet transiting LHS 475, an M3 dwarf located 12.5 pc from the Sun. The mass of the star is $0.274 \pm 0.015$ \msun. The planet, originally reported as TOI 910.01, has an orbital period of $2.0291025 \pm 0.0000020$ days and an estimated radius of $0.955 \pm 0.053$ \rearth. We confirm the validity and source of the transit signal with MEarth ground-based follow-up photometry of five individual transits. We present radial velocity data from CHIRON that rule out massive companions. In accordance with the observed mass-radius distribution of exoplanets as well as planet formation theory, we expect this Venus-sized companion to be terrestrial, with an estimated RV semi-amplitude close to 1.0 m/s. LHS 475 b is likely too hot to be habitable but is a suitable candidate for emission and transmission spectroscopy.

\end{abstract}

%% Keywords should appear after the \end{abstract} command. 
%% The AAS Journals now uses Unified Astronomy Thesaurus concepts:
%% https://astrothesaurus.org
%% You will be asked to selected these concepts during the submission process
%% but this old "keyword" functionality is maintained in case authors want
%% to include these concepts in their preprints.
%% \keywords{TBD}

\section{Introduction} \label{sec:intro}

The Transiting Exoplanet Survey Satellite \citep[TESS;][]{Ricker2015} was launched in 2018 and has already yielded a plethora of planet discoveries. At the time of this writing, there are 326 confirmed TESS planets as well as thousands of TESS Objects of Interest (TOIs) still awaiting validation. Of particular interest are planets transiting nearby red dwarfs due to the relatively larger size of the planet compared to the star. This leads to a larger transit depth, facilitating planet detection. It also increases the feasibility for detailed atmospheric characterization of the planet via transit spectroscopy, which is among the goals of the recently launched James Webb Space Telescope (JWST). Only a handful of stars with masses below 0.3 \msun~located within 15 pc of the Sun are known to host transiting planets: GJ 1132 \citep{BertaThompson2015}, GJ 1214 \citep{Charbonneau2009}, LHS 1140 \citep{Dittmann2017,Ment2019}, LHS 3844 \citep{Vanderspek2019}, LTT 1445A \citep{Winters2019,Winters2022}, TOI 540 \citep{Ment2021}, and TRAPPIST-1 \citep{Gillon2016,Gillon2017}.

Follow-up observations of TOIs, typically by ground-based observatories, are essential for multiple reasons. Firstly, they provide an independent validation of the transiting planet, ruling out instrument systematics as the cause of the transit-like signal. Secondly, the sky-projected size of each individual TESS pixel is large ($21\arcsec \times 21\arcsec$) and typically contains many background light sources in addition to the intended target. Subsequently, follow-up observations are crucial to exclude background stars (including unresolved eclipsing binaries) as potential sources of the transit signal. Thirdly, ground-based photometric observations are a relatively cost-effective way to increase the total number of observed transits, leading to more precise estimates for modeled system parameters as well as refined ephemerides, which are often essential for planning additional follow-up observations. Finally, spectroscopic follow-up observations can yield estimates for the planet's mass, a crucial parameter that cannot be determined from transit photometry.

The vast majority of confirmed transiting planets have estimated radii larger than that of the Earth. Only six sub-Earth-sized planets are currently known to orbit stars within a distance of 15 pc from the Sun, including GJ 367 b \citep{Lam2021}, L 98-59 b \citep{Kostov2019}, TOI 540 b \citep{Ment2021}, and TRAPPIST-1 d, e, and h \citep{Gillon2016,Gillon2017,Agol2021}. Unlike planets substantially larger than Earth, sub-Earth-sized planets are likely too small to hold on to a substantial volatile envelope, and therefore they are highly likely to be terrestrial in nature with a mostly rocky interior composition. The existence of a population of terrestrial planets distinct from larger sub-Neptunes and water worlds is evidenced by the bimodal radius distribution of planets \citep{Owen2013,Fulton2017}. The gap between the two populations occurs at planet radii of 1.6-2.0 \rearth~for Sun-like stars \citep{Fulton2018} and moves down to 1.5 \rearth~for M dwarfs \citep{Cloutier2020}, with planets below these radii mainly expected to have terrestrial bulk compositions \citep{Weiss2014,Dressing2015,Rogers2015}. Therefore, we expect a sub-Earth-sized planet to be terrestrial with a high degree of confidence even in the absence of a mass estimate from radial velocity or transit timing variation data.

Here we present the discovery and subsequent ground-based validation observations of LHS 475 b, a Venus-sized planet orbiting a nearby M dwarf. Section \ref{sec:star} describes the properties of the host star. Section \ref{sec:data} summarizes various data sets used in this study. We describe our modeling procedure in Section \ref{sec:modeling} and present results in Section \ref{sec:results}. Finally, Section \ref{sec:discussion} summarizes the article and offers a brief discussion.

\section{Stellar Parameters of LHS 475} \label{sec:star}

LHS 475 (also known as 2MASS J19205439-8233170 and L 22-69) is a main-sequence red dwarf belonging to the M3 spectral class \citep{Hawley1996}. Based on the trigonometric parallax of $\pi = 80.113 \pm 0.021$ mas from Gaia Data Release 3 \citep{Gaia2016,GaiaDR3}, we calculate a distance of $12.482 \pm 0.003$ pc. Gaia DR3 determined its proper motion to be $\mu_\alpha = 342.30 \pm 0.03$ mas yr$^{-1}$ and $\mu_\delta = -1230.30 \pm 0.02$ mas yr$^{-1}$.

There are no additional light sources listed within 10" in either the 2MASS catalog or the TESS Input Catalog (TIC). Leveraging the high proper motion of the target, we used archival images from the Two Micron All-Sky Survey \citep[2MASS;][]{Skrutskie2006} as well as the Digitized Sky Survery (DSS) to rule out the presence of any distant background stars at the location of LHS 475 at the time of the observations presented in this paper (years 2019-2021). The 2MASS images were taken in year 2000 while the DSS images are from years 1976-1999. While there are no visible sources in the 2MASS J, H, and K images, the bluer DSS images do reveal an additional light source at a sky-projected distance of 7" from LHS 475 at the time of the observations. We later identified this background source within the Gaia DR3 catalog with an ID of 6347643496607834880. It has a $B_P - R_P$ color of 1.05 mag, in contrast to the $B_P - R_P = 2.73$ mag estimate for LHS 475. It is 6.7 magnitudes fainter than LHS 475 in the Gaia bandpass, and based on the color difference, we estimate a difference close to 8 magnitudes in the TESS and MEarth bandpasses. As a result, this object would be too faint to produce the planetary transits described within this work. We further investigate the presence of other nearby bright objects in Section \ref{sec:speckle}.

We adopt a $K$-band apparent brightness of $K = 7.686 \pm 0.042$ mag from 2MASS, yielding an absolute brightness of $M_K = 7.205 \pm 0.042$ mag. We also adopt $J$ and $H$-band fluxes from 2MASS and $VRI$-photometry from \citet{Jao2011}. Using the mass-luminosity relation for main-sequence M dwarfs in \citet{Benedict2016}, we estimate a stellar mass of $M = 0.274 \pm 0.015$ \msun. We use two independent radius-mass relations to estimate the stellar radius: we obtain $R = 0.281 \pm 0.013$ \rsun~from optical interferometry of single stars \citep{Boyajian2012} and $R = 0.291 \pm 0.014$ \rsun~from eclipsing binary measurements \citep{Bayless2006}. We combine the two and adopt a weighted average of $R = 0.286 \pm 0.010$ \rsun~as the final radius. We check for consistency with the mass-radius relation for M dwarfs by \citet{Mann2015} which yields $R = 0.287 \pm 0.026$ \rsun.

We employ bolometric corrections (BC) to determine the luminosity of LHS 475. We use Table 5 of \citet{Pecaut2013} to interpolate between the $V-K$ color and $BC_V$, obtaining a $V$-band correction value of $BC_V = -2.35$ mag and a bolometric luminosity of $L = 0.00907$ \lsun. Next, we apply the third-order polynomial fit between $V-J$ and $BC_V$ in \citet[and subsequent erratum]{Mann2015} to derive $BC_V = -2.25$ and $L = 0.00829$ \lsun. Finally, the relationship between $BC_K$ and $I-K$ in \citet{Leggett2001} produces $BC_K = 2.70$ and $L = 0.00870$ \lsun. We adopt as the stellar luminosity the average of the three estimates; therefore, $L = 0.00869 \pm 0.00039$ \lsun. The uncertainty in $L$ was taken from the unbiased sample variance of the individual estimates. We then proceed to use the Stefan-Boltzmann law to determine the effective stellar surface temperature, obtaining $\teff = 3295 \pm 68$ K. This result was based on the solar values of $M_{\rm bol, \sun} = 4.7554$ mag and $T_{\rm eff, \sun} = 5772$ K published by \citet{Mamajek2012}.

LHS 475 appears to be a magnetically quiet star. The CHIRON spectra (described in Section \ref{sec:chiron}) do not show evidence of rotational broadening, and H$\alpha$ is in absorption. LHS 475 has a previously determined photometric rotation period of $\prot = 79.3$ days \citep{Newton2018}, consistent with its relative inactivity. It also flares relatively infrequently; specifically, the star has an estimated flare rate of $\ln R_{31,5} = -11.90 \pm 0.79$ where $R_{31,5}$ is the number of flares per day with total energy above $3.16 \times 10^{31}$ ergs in the TESS bandpass \citep{Medina2020}.

While M dwarfs do slowly spin down as they evolve, accurately determining their age is known to be a difficult endeavor. For the slowest-rotating mid-to-late M dwarfs ($\prot > 70$ days), galactic kinematics suggests a mean age of $5^{+4}_{-2}$ Gyr \citep{Newton2016}. More recently, \citet{Medina2022} used galactic kinematics to derive a mean age of $5.6 \pm 2.7$ Gyr for mid-to-late M dwarfs with rotation periods between 10-90 days while such stars with $\prot > 90$ days would have an estimated age of $12.9 \pm 3.5$ Gyr. Ultimately, we are unable to ascertain the age of LHS 475, but it is unlikely to be less than a few billion years old.

\section{Data} \label{sec:data}

\subsection{TESS} \label{sec:tess}

The Transiting Exoplanet Survey Satellite (TESS) gathered photometric measurements of LHS 475 during its Prime Mission (observation sectors 12 and 13) as well as its first Extended Mission (sectors 27 and 39). These observations span a Barycentric Julian Date (BJD) range of 2458625.0-2459389.7 between May 2019 and June 2021. The target was included in the TESS Input Catalog (TIC) with a TIC ID of 369327947 as well as the original TESS Candidate Target List \citep[CTL;][]{Stassun2018}. LHS 475 was also included in TESS Guest Investigator Programs G011180 (PI: Courtney Dressing) and G011231 (PI: Jennifer Winters). The observations were made with a two-minute cadence during the Prime Mission and a 20-second cadence within the Extended Mission. We utilize photometric data reduced by the NASA Ames Science Processing Operations Center (SPOC) pipeline \citep{Jenkins2016}. Specifically, we adopt the Pre-search Data Conditioning Simple Aperture Photometry \citep[PDCSAP;][]{Smith2012,Stumpe2012,Stumpe2014} two-minute cadence version of the light curve which has been cotrended and corrected for instrument systematics as well as crowding: unresolved light from other sources that fall within the same TESS CCD pixel. We exclude 6 measurements that were marked through the SPOC Data Quality Flags to correspond to impulsive outliers (cadence quality flag bit 10). We also manually exclude a section of the data at the end of sector 27 spanning the BJD range 2459059.5-2459060.2 (comprising 0.68\% of the total number of data points) due to a strong negative trend in the observed flux that is indicative of spacecraft systematics. The remaining data set contains a total of 73,604 individual measurements.

SPOC initially detected a transit-like signal in phase-folded sector 12 data and the TESS team dubbed the planet candidate TESS Object of Interest (TOI) 910.01. The accompanying Data Validation Report \citep[DVR;][]{Twicken2018,Li2019} cited a candidate period of 2.029 days, a transit depth close to 1100 parts per million (ppm), and a transit signal-to-noise ratio (SNR) of 12.0 based on 12 transits from sector 12. The detected signal passed preliminary validation tests and had a very low bootstrap false alarm probability, favoring the hypothesis that TOI 910.01 indeed represents a planet. An updated DVR from SPOC based on all four observation sectors later updated the orbital period to $P = 2.029088 \pm 0.000006$ days, the transit depth to $978 \pm 73$ ppm, and the SNR to 19.3 based on a total of 45 transits. We note that despite the availability of 20-second cadence data from the Extended Mission, SPOC only performs a transit search in the 2-minute cadence version of those data. Either cadence is much shorter than the fitted transit duration of 41.6 minutes.

\begin{figure}
    \includegraphics[width=\textwidth]{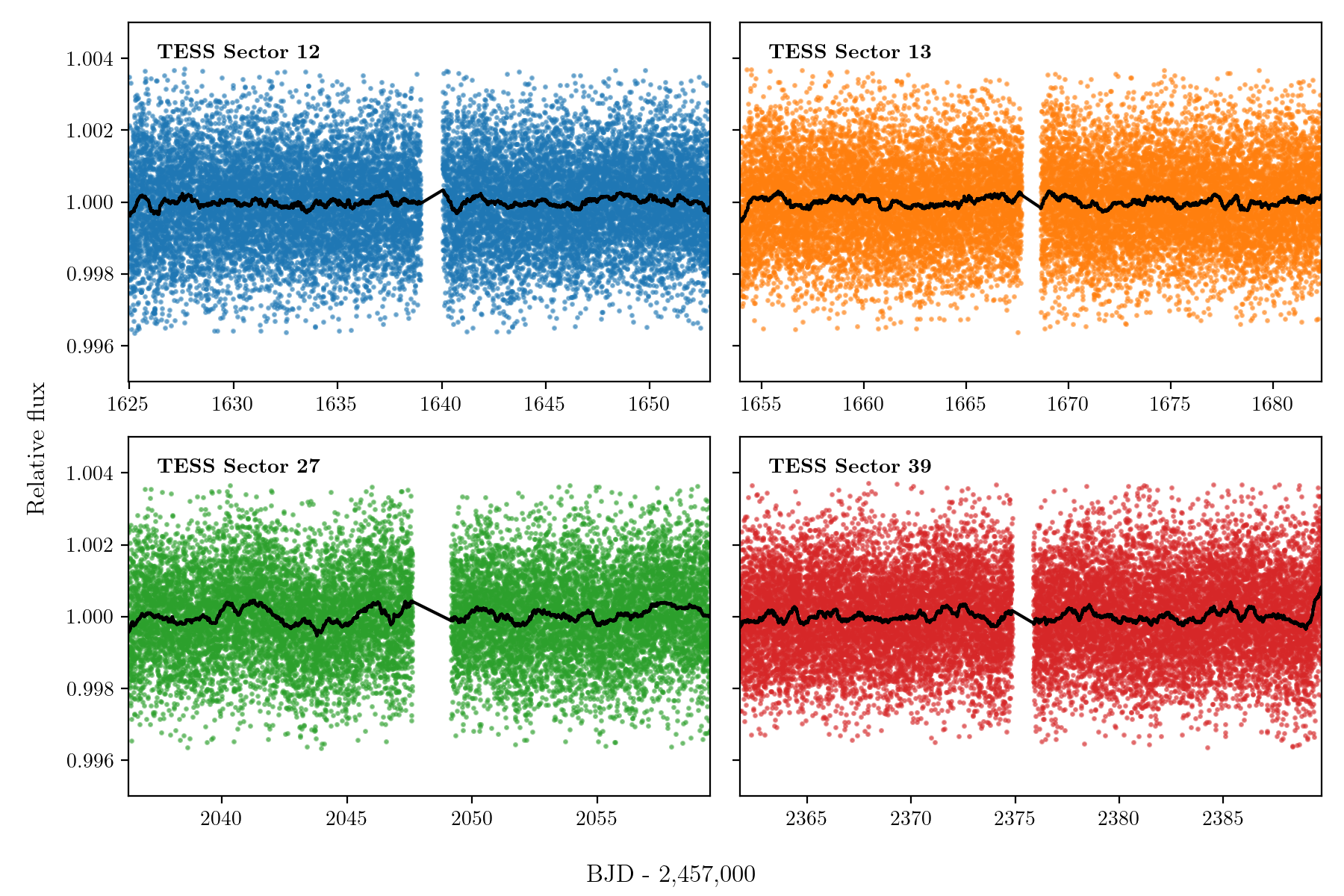}
    \caption{TESS photometry of LHS 475 in Sectors 12, 13, 27, and 39. The black line represents the fitted baseline as described in Section \ref{sec:tess}.}
    \label{fig:tess}
\end{figure}

Since the estimated rotation period of LHS 475 ($\prot = 79.3$ days) is substantially longer than the duration of a single TESS sector, we do not attempt to include an explicit model for the quasi-periodic flux variation caused by stellar rotation. Instead, we de-trend the light curve using a sliding median with a fixed width of 12 hours. This approach has the benefit of smoothing out any long-term variations while not affecting the light curve on short timescales (e.g. the estimated duration of a transit). We provide a plot of the TESS light curve as well as the fitted baseline within each observation sector in Figure \ref{fig:tess}. All the TESS data used in this paper can be found in MAST: \dataset[10.17909/bmdh-kd60]{http://dx.doi.org/10.17909/bmdh-kd60}.

\subsection{MEarth} \label{sec:mearth}

We obtained follow-up observations for five individual transits of LHS 475 b with the ground-based MEarth-South telescope array \citep{Nutzman2008,Berta2012,Irwin2015}. The MEarth-South array consisted of eight robotically controlled 40-centimeter telescopes at the Cerro Tololo Inter-American Observatory in Chile. Each telescope was equipped with a CCD camera that is sensitive to red optical and near-infrared light. The observations took place in August and September of 2019; the corresponding Barycentric Julian Date (BJD) range is 2458717.5-2458727.8. A total of 7 individual telescopes were used concurrently to monitor the star during transit events, and a separate light curve was produced for each telescope as a result of the data reduction process.

We perform aperture photometry with fixed aperture radii of 12, 17, 24, and 34 pixels (at a scale of 0.84\arcsec~per pixel) and verify that the transit signal persists regardless of the choice of aperture. We also generate light curves for every detected light source within 2.5\arcmin~of LHS 475 and verify that the transit signal is not present in any of them. We adopt an aperture radius of 12 pixels (or 10.08\arcsec) for the remainder of the analysis.

\begin{figure}
    \includegraphics[width=0.7\textwidth]{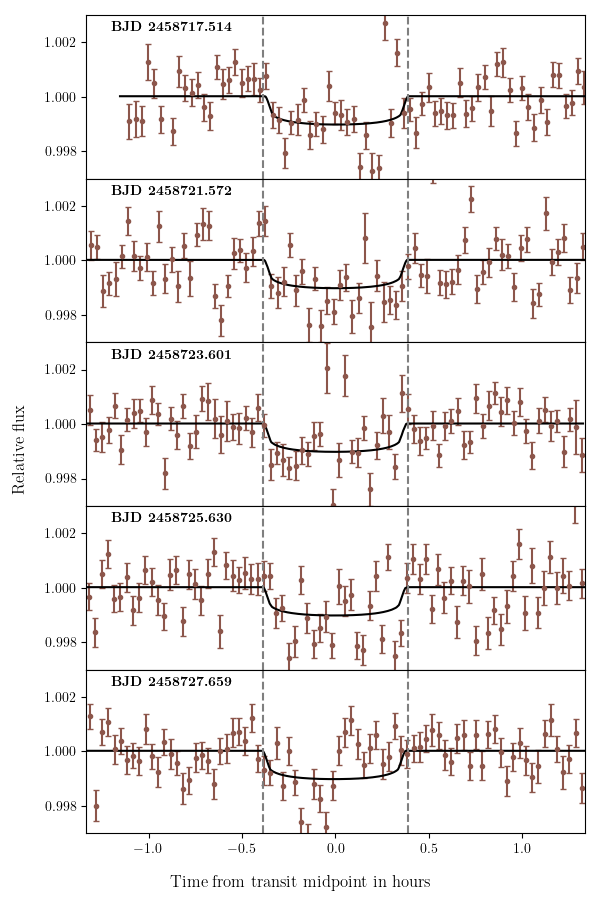}
    \centering
    \caption{MEarth-South follow-up photometry of five individual transits of LHS 475. The Barycentric Julian Date values (BJD) within the plots denote the times of transit midpoints. The vertical dashed lines encompass the full transit duration. The solid line depicts the best-fit model for all MEarth data; it is identical to the model in the middle panel of Figure \ref{fig:combined} and is further described in Section \ref{sec:results_mearth}.}
    \label{fig:mearth}
\end{figure}

The MEarth light curves need to be detrended to account for atmospheric effects and telescope systematics. We begin by estimating a full transit duration of $T_{\rm full} = 40.0$ minutes based on the TESS data. For each observed transit, we fit a baseline to the MEarth light curve, including all points within $2\tfull$ of the transit midpoint $t_0$ while excluding the transit window itself; that is, we include every observation timestamp $t$ such that
\begin{displaymath}
    0.5 \leq \left|\frac{t - t_0}{\tfull}\right| \leq 2
\end{displaymath}
In order to best describe the behavior of each transit's baseline, we utilize several competing models. Specifically, we fit each baseline with a constant, linear, and quadratic function. We adopt the model with the lowest value for the Bayesian information criterion (BIC) which accounts for the number of model parameters. We display the detrended transit light curves in Figure \ref{fig:mearth}. The baselines for the first two of the five transits were fit with a linear function while the remaining three were detrended using a quadratic function.

LHS 475 was monitored as part of the MEarth survey from October 2016 until August 2018. Over that time, the star accumulated 12,836 photometric observations. However, LHS 475 b was not detected by the MEarth team due to the small transit depth induced by the planet (approx. 1 ppt) and the relatively low cadence of observations. While the planet remained undetected, this data set enabled the critical photometric determination of the stellar rotation period of 79.317 days by \citet{Newton2018}.

\subsection{CHIRON} \label{sec:chiron}

\begin{figure}
    \includegraphics[width=\textwidth]{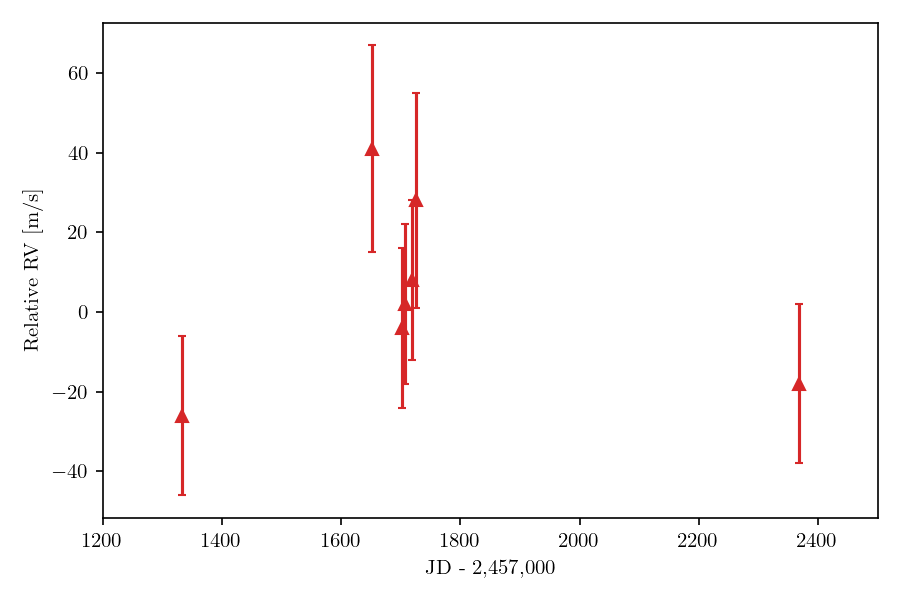}
    \caption{Relative radial velocities of LHS 475 from CHIRON. The estimated systemic velocity is $-10.3 \pm 0.5$ km/s.}
    \label{fig:rv}
\end{figure}

We utilized the CHIRON spectrograph \citep{Tokovinin2013} mounted on the 1.5-meter SMARTS telescope at the Cerro Tololo Inter-American Observatory (CTIO) to gather 7 reconnaissance spectra of LHS 475. Four of these spectra were obtained as part of the volume-complete spectroscopic survey of nearby mid-to-late M dwarfs by \citet{Winters2021}. We carried out the observations between August 2018 and June 2021. We produced radial velocities (RVs) from each of the spectra according to the reduction methods described in detail in Pass et al. (2023, submitted). The reduction process involves performing a cross-correlation between the CHIRON spectra and a set of rotationally broadened templates of inactive stars over wavelength ranges in the regime of 6400-7850 \AA. We provide a list of relative RVs in Table \ref{tbl:rv} and display the entire RV time series in Figure \ref{fig:rv}. Note that the quoted RVs do not include a derived systemic RV of $-10.3 \pm 0.5$ km/s, where the RV uncertainty is dominated by the cross-correlation process. We also adopt a noise floor of 20 m/s based on instrument instability for mid-to-late M dwarfs (Pass et al. 2023, submitted).

The RV time series does not show any evidence of an overall trend, which would point to the existence of a massive companion. The RMS of the RVs is 22.1 m/s, very close to the quoted individual RV uncertainties. In addition, we are able to rule out companions more massive than 0.20 Jupiter masses at the orbital period of the planet with 99.73\% ($3\sigma$) confidence. We obtained this value by fitting an RV model with the orbital period, eccentricity, and time of mid-transit fixed to the values given in Table \ref{tbl:results} and calculating the required RV semi-amplitude such that the $\chi^2$ values of the fits, each with $N_{\rm{RV}} - 1$ degrees of freedom, have associated p-values below 0.0027.

The spectra show H$\alpha$ in absorption. \citet{Medina2020} obtained an H$\alpha$ equivalent width of $0.26 \pm 0.09$ \AA~using a subset of these spectra.

% Table of RVs
\begin{deluxetable}{lrr}
    \tabletypesize{\footnotesize}
	\tablewidth{0pt}
	\tablecaption{CHIRON relative RVs of LHS 475\label{tbl:rv}}
	\tablehead{
		BJD (TDB) & RV & Uncertainty\\
		 & (m/s) & (m/s)
	}
	\startdata
    2458332.6635 & -26 & 20\\
	2458650.7860 & 41 & 26\\
    2458701.7103 & -4 & 20\\
    2458707.6332 & 2 & 20\\
    2458718.6361 & 8 & 20\\
    2458724.5771 & 28 & 27\\
    2459367.7897 & -18 & 20\\
	\enddata
\end{deluxetable}

\subsection{Zorro}\label{sec:speckle}

\begin{figure}
    \includegraphics[width=\textwidth]{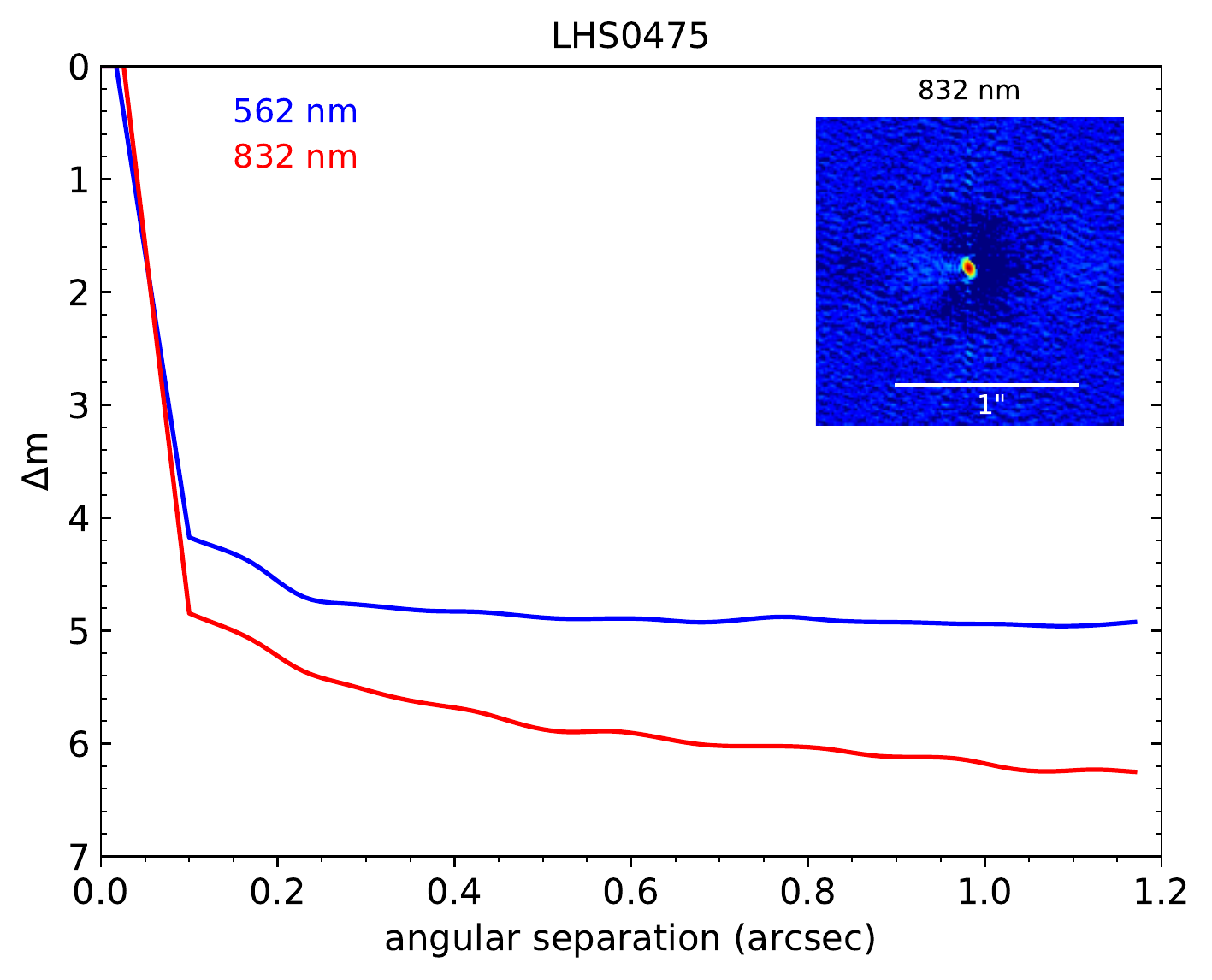}
    \caption{Speckle imaging sensitivity curves of LHS 475 utilizing the 562 nm (blue line) and 832 nm (red line) bandpasses. No nearby light sources are detected.}
    \label{fig:speckle}
\end{figure}

We performed speckle imaging observations of LHS 475 with the Zorro dual-channel imager on the 8.1-meter Gemini South telescope in Chile (program number GS-2019A-Q-302, PI: Winters). The star was observed on 18 July 2019 in the 562 nm and 832 nm bandpasses. The images are diffraction-limited with an estimated FWHM of 0.02\arcsec. We provide the detection sensitivity curve and the auto-correlation function of the imaging data in Figure \ref{fig:speckle}. No nearby light sources (within 1.2\arcsec, corresponding to 15 AU) are detected in either band down to an estimated contrast $\Delta m$ of 4.88 mag (562 nm) or 5.87 mag (832 nm) at 0.5\arcsec~from the star.

\section{Modeling} \label{sec:modeling}

We utilize the \exoplanet~package \citep{exoplanet:joss,exoplanet:zenodo} in Python for transit modeling purposes. \exoplanet~employs the modeling framework of the PyMC Python library \citep{exoplanet:pymc3} and uses \textsf{Theano}~\citep{exoplanet:theano} for computationally efficient Markov Chain Monte Carlo sampling. Limb-darkened transit signals are computed analytically by the open-source \textsf{starry}~package \citep{exoplanet:luger18}.

We adopt an identical light curve and transit model for TESS and MEarth data sets, constraining the modeled parameters with various prior probability distributions. Specifically, the stellar mass and radius are constrained with Gaussian priors; in each case, the location and width of the prior matches the estimated value from Section \ref{sec:star}. Gaussian priors are also included for the orbital period $P$ and time of mid-transit $t_0$, although the posterior distributions for either parameter constrain these values much more tightly. We impose a uniform prior on the planet-to-star radius ratio with a domain of $r/R \in [0, 0.1]$. Conditioned on the value of $r/R$, the prior distribution for the impact parameter $b$ is distributed uniformly between 0 and $1+r/R$. We also include a loose Gaussian prior centered at a value of 1 on the baseline relative flux of the light curve.

We opt for a quadratic limb darkening model for the stellar surface. The limb darkening coefficients $(u_1, u_2)$ are taken from Table 15 of \citet{Claret2018} based on the spherical PHOENIX-COND model \citep{Husser2013}. Specifically, we adopt $u_1 = 0.1529$ and $u_2 = 0.4604$ which correspond to a surface gravity of $\log g = 5.0$ and a surface temperature of 3300 K. The metallicity is implicitly assumed to be equal to the Solar value. While these coefficients were computed for the TESS bandpass, the MEarth spectral response is sufficiently similar to the one employed by TESS such that we have not found meaningful differences in the estimated limb darkening coefficients in previous works \citep[e.g.][]{Ment2021}.

The eccentricity $e$ was fixed to a value of 0 as planets orbiting this close to their stars are expected to be tidally circularized. In order to estimate the tidal circularization timescale $\tau_e$, we utilize a simplified equation that \citet{Rasio1996} adopted based on the work by \citet{Goldreich1966}:
\begin{displaymath}
    \frac{1}{\tau_e} \equiv -\frac{1}{e}\frac{de}{dt} \approx \frac{63}{4Q} \sqrt{\frac{GM}{a^3}} \left(\frac{M}{m}\right) \left(\frac{r}{a}\right)^5
\end{displaymath}
where $Q$ is the specific dissipation parameter, $m$ and $r$ are the planet's mass and radius, $M$ is the stellar mass, and $a$ is the semi-major axis. Assuming an Earth-mass planet and $Q < 500$ (an upper limit based on terrestrial Solar System planets and satellites), we find $\tau_e < 9$ Myr, much less than the probable age of the system. Thus, we expect the orbit to be close to circular.

We perform separate model optimization and sampling runs for three different data sets: (1) TESS photometry, (2) MEarth photometry, and (3) a combined set of TESS and MEarth data. The initial optimization is carried out by \pymc~using a limited memory Broyden-Fletcher-Goldfarb-Shanno (BFGS) algorithm and it returns a local maximum a posteriori (MAP) estimate. We then draw 1000 samples from the posterior, starting at the MAP point, after tuning for 500 iterations before sampling. We utilize multiprocess sampling with 2 chains, resulting in a total of 2000 drawn samples. We then adopt the mean and standard deviation from the marginal distributions of each variable. We present the results of the modeling in the following section.

\section{Results} \label{sec:results}

\subsection{TESS only}\label{sec:results_tess}

\begin{figure}
    \includegraphics[width=\textwidth]{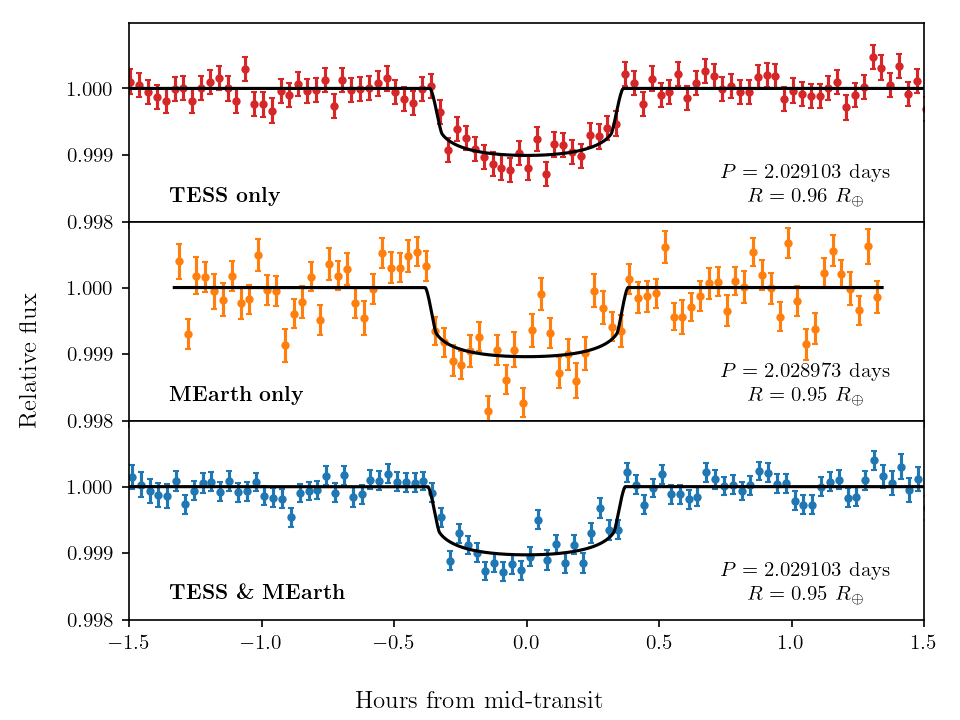}
    \caption{Phase-folded photometry of LHS 475 highlighting the 3-hour window surrounding the planetary transits. Solid lines depict a best-fit model using the \textit{maximum a posteriori} (MAP) estimates of the orbital period $P$ and planet radius $R$. We produce separate subplots for TESS data (top panel), MEarth data (middle panel), and the combined TESS \& MEarth data set (bottom panel).}
    \label{fig:combined}
\end{figure}

Fitting the model with the four sectors of TESS data alone yields a tightly constrained ephemeris with an orbital period of $P = 2.0291019 \pm 0.0000025$ days and a transit mid-point BJD of $t_0 = 2,458,626.20451 \pm 0.00045$. The orbital period is roughly 2$\sigma$ away from the SPOC estimate quoted in Section \ref{sec:tess}. Apart from differences in modeling and sampling, this discrepancy could be affected by the data selection process: the SPOC analysis excluded chunks of data at the beginning and end of each sector due to spurious flux trendlines caused by instrumental systematics. We successfully de-trended most of these regions. We obtain a fitted planetary radius of $r = 0.955 \pm 0.056$ \rearth~and an impact parameter of $b = 0.731 \pm 0.035$. We plot a phase-folded version of our best-fit model utilizing TESS data in the top panel of Figure \ref{fig:combined}.

\subsection{MEarth only}\label{sec:results_mearth}

Repeating the modeling and sampling process with the MEarth data set (consisting of follow-up light curves of five individual transits), we estimate an orbital period of $P = 2.02894 \pm 0.00015$ days and a transit mid-point BJD of $t_0 = 2,458,626.2104 \pm 0.0065$. These values are consistent within 1$\sigma$ with the TESS-based results in the previous section. We calculate a planet radius of $r = 0.990 \pm 0.070$ \rearth; however, the best-fit \textit{maximum a posteriori} (MAP) estimate is 0.95\rearth. Both of these values are consistent with TESS data. The fitted impact parameter from MEarth data is $b = 0.753 \pm 0.064$. The phase-folded model based on the MAP estimate is displayed in the middle panel of Figure \ref{fig:combined}.

\subsection{Combined TESS \& MEarth}

Noting that the calculated parameter values yielded by the two data sets are remarkably consistent, we optimize the model again for the combined TESS \& MEarth light curves in order to produce final estimates for the model parameters. This produces an orbital period of $P = 2.0291025 \pm 0.0000020$ days and a transit mid-point BJD of $t_0 = 2,458,626.20421 \pm 0.00029$. We find a planetary radius of $r = 0.955 \pm 0.053$ \rearth~and an impact parameter of $b = 0.705 \pm 0.037$, corresponding to an inclination angle of $i = 87.44 \pm 0.27$ degrees. These values are consistent with previous estimates, with the uncertainties in the transit ephemeris largely set by the TESS light curve due to the substantially larger number of observed transits and a longer time baseline within that data set. The best-fit combined model is displayed in the bottom panel of Figure \ref{fig:combined}. We adopt the values from the combined analysis as final estimates for LHS 475 b. The values for the system parameters are summarized in Table \ref{tbl:results}.

% Table of results
\begin{deluxetable}{lcc}
    \tabletypesize{\footnotesize}
	\tablewidth{0pt}
	\tablecaption{System parameters for LHS 475\label{tbl:results}}
	\tablehead{
		Parameter & Values for LHS 475 & Source\tablenotemark{a}
	}
	\startdata
	\multicolumn{3}{l}{\textbf{Stellar parameters}}\\
	Right ascension (J2000) & 19h 20min 54.38s & (1)\\
	Declination (J2000) & -82$^\circ$ 33' 16.17" & (1)\\
	\multirow{2}{*}{Proper motion (mas yr$^{-1}$)} & 
	    $\mu_\alpha = 342.30 \pm 0.03$ & \multirow{2}{*}{(1)}\\
	    & $\mu_\delta = -1230.30 \pm 0.02$ & \\
	\multirow{6}{*}{Apparent brightness (mag)} &
	        $V = 12.69 \pm 0.03$ & (2)\\
	        & $R = 11.51 \pm 0.03$ & (2)\\
	        & $I = 10.00 \pm 0.03$ & (2)\\
	        & $J = 8.555 \pm 0.030$ & (3)\\
	        & $H = 8.004 \pm 0.038$ & (3)\\
	        & $K = 7.686 \pm 0.042$ & (3)\\
	Distance (pc) & 12.482 $\pm$ 0.003 & (1)\\
	Mass (\msun) & 0.274 $\pm$ 0.015 & (4)\\
	Radius (\rsun) & 0.286 $\pm$ 0.010 & (4)\\
	Luminosity (\lsun) & 0.00869 $\pm$ 0.00039 & (4)\\
	Effective temperature (K) & 3295 $\pm$ 68 & (4)\\
	Rotational period (days) & 79.317 & (5)\\
	\hline\hline
	Parameter & Values for LHS 475 b & \\
	\hline
	\multicolumn{3}{l}{\textbf{Modeled transit parameters}}\\
	Orbital period $P$ (days) & 2.0291025 $\pm$ 0.0000020 & \\
	Eccentricity $e$ & 0 (fixed) & \\
	Time of mid-transit $t_0$ (BJD) & 2458626.20421 $\pm$ 0.00029 & \\
	Impact parameter $b$ & 0.705 $\pm$ 0.037 & \\
	Planet-to-star radius ratio $r/R$ & 0.03124 $\pm$ 0.00065 & \\
	$a/R$ ratio & 15.77 $\pm$ 0.77 & \\
	\hline
	\multicolumn{3}{l}{\textbf{Derived planetary parameters}}\\
	Radius $r$ (\rearth) & 0.955 $\pm$ 0.053 & \\
	Semi-major axis $a$ (AU) & 0.02042 $\pm$ 0.00036 & \\
	Inclination $i$ (deg) & 87.44 $\pm$ 0.27 & \\
	Bolometric incident flux $S$ (\searth) & 20.8 $\pm$ 1.1 & \\
	Equilibrium temperature\tablenotemark{b} $T_{\rm eq}$ (K) & 587 $\pm$ 18 & \\
	\enddata
	\tablenotetext{a}{(1) Gaia DR3, (2) \citet{Jao2011}, (3) 2MASS, (4) This work, (5) \citet{Newton2018}.}
	\tablenotetext{b}{The equilibrium temperature assumes a Bond albedo of 0. For an albedo of $A$, the reported temperature has to be multiplied by $(1-A)^{1/4}$.}
\end{deluxetable}

\section{Discussion and conclusion} \label{sec:discussion}

LHS 475 b is a Venus-sized planet orbiting a nearby magnetically quiet M dwarf. Its host star has an estimated luminosity equal to 0.9\% of the Solar value and it is located at a distance of 12.5 pc from the Sun. We determine an orbital period of $P = 2.0291025 \pm 0.0000020$ days (corresponding to an $a/R$-ratio of $15.77 \pm 0.77$) which is short enough for the planet to be tidally locked with a high degree of confidence\footnote{We use Eq. 3 of \citet{Pierrehumbert2019}, based on the results of \citet{Goldreich1966}, to estimate a tidal locking timescale of 290 years for LHS 475 b.}. Thus, the planet is likely to have an uneven surface temperature, with the exact temperature profile dependent on the composition and dynamics of the planetary surface and atmosphere. However, we can make a first-order estimate by assuming that the planet absorbs all incoming stellar radiation hitting its cross-section and produces black body radiation uniformly from its entire surface according to the Stefan-Boltzmann law. The estimated incident bolometric flux at the planet's orbit is $S = 20.8 \pm 1.1$ \searth. Balancing the input and output energy fluxes yields an estimated equilibrium temperature of $\teq = 587 \pm 18$ K for a zero-albedo surface, $\teq = 537 \pm 16$ K for an Earth-like Bond albedo ($A_B = 0.3$), and $\teq = 407 \pm 12$ K for a Venusian Bond albedo ($A_B = 0.77$). All of these values are too high for the planet to be habitable in the traditional sense. Furthermore, if the absorbed incident radiation is emitted from the dayside only (as would be the case with tidal locking and zero heat redistribution), the dayside temperatures would be 19\% higher than the values quoted above.

However, hot effective surface temperatures make the planet more amenable to characterization via emission and transmission spectroscopy during its transits. Adopting the \citet{Kempton2018} framework for a zero-albedo model with full day-night heat redistribution, we obtain a transmission spectroscopic metric (TSM) value of 27.7 (assuming the appropriate scale factor of 0.19). TSM is proportional to the expected transmission spectroscopy S/N. This quoted value places LHS 475 b among a group of other nearby small planets that are promising targets for transit spectroscopy, such as TRAPPIST-1 c \citep[estimated TSM of 24.1;][]{Gillon2017,Agol2021} and LHS 1140 c \citep[$\rm{TSM} = 25.4$;][]{Ment2019}. We note that transmission spectroscopy observations of LHS 475 b were recently conducted with the James Webb Space Telescope during two planetary transits \citep[under review]{LustigYaeger2023}. We can similarly calculate the emission spectroscopic metric (ESM) and find that $\rm{ESM} = 5.2$ for LHS 475 b, comparable to LHS 1445A b \citep[$\rm{ESM} = 5.7$;][]{Winters2019,Winters2022}.

Due to its tidal locking, LHS 475 b is also a feasible target for photometric thermal emission measurements with JWST during a secondary eclipse. Such observations can quantify the amount of heat redistibution between the dayside and the nightside, which can be used to constrain the presence of an optically thick planetary atmosphere. Among the known terrestrial planets orbiting nearby stars, LHS 3844 b is the only planet where a thick atmosphere has been ruled out \citep{Kreidberg2019,DiamondLowe2020}. The surface insolation of LHS 475 b is lower than that of LHS 3844 b, making the former more likely to retain an atmosphere. If an atmosphere were to be found on LHS 475 b, it would place an important constraint on atmospheric escape in terrestrial planets. Furthermore, we note that LHS 475 b has a very similar radius and surface insolation to TOI 540 b \citep{Ment2021}; however, the latter is orbiting an M dwarf that is still within the magnetically active phase of its evolution. Consequently, a comparative study of the two planets' atmospheres could yield an important before-and-after test for atmospheric escape.

Our best estimate for the radius of LHS 475 b is $r = 0.955 \pm 0.053$ \rearth. A similar value can be independently derived from either TESS photometry (see Section \ref{sec:results_tess}) or MEarth photometry (Section \ref{sec:results_mearth}) alone with remarkable consistency, demonstrating the power of targeted ground-based follow-up observations to confirm the validity of TESS planet candidates. Unfortunately, the existing radial velocity data of LHS 475 does not have the necessary precision to calculate the mass of the planet. As explained in Section \ref{sec:intro}, planets of this size are highly likely to be terrestrial. Given that the planet is close in size to Earth, it is not unreasonable to suppose that it might also have a similar interior composition. Adopting a simple two-layer composition model with an Earth-like core mass fraction (CMF) of 0.33, we invert the empirical radius-mass relation by \citet{Zeng2016} and derive a planetary mass of 0.84 \mearth.\footnote{Adopting a Venusian CMF of 0.31 yields a similar mass estimate of 0.83 \mearth.} Subsequently, we estimate that the radial velocity signal induced by LHS 475 b will have a semi-amplitude of 1.0 m/s. This level of precision is achievable with current state-of-the-art RV instruments but would likely require a substantial amount of observation time.

\begin{acknowledgments}
The MEarth team acknowledges funding from the David and Lucile Packard Fellowship for Science and Engineering (awarded to DC). This material is based on work supported by the National Science Foundation under grants AST-0807690, AST-1109468, AST-1004488 (Alan T. Waterman Award) and AST-1616624. This publication was made possible through the support of a grant from the John Templeton Foundation. The opinions expressed in this publication are those of the authors and do not necessarily reflect the views of the John Templeton Foundation. This material is based upon work supported by the National Aeronautics and Space Administration under Grant No. 80NSSC18K0476 issued through the XRP Program. Funding for the TESS mission is provided by NASA's Science Mission Directorate. We acknowledge the use of public TESS data from pipelines at the TESS Science Office and at the TESS Science Processing Operations Center. This paper includes data collected by the TESS mission that are publicly available from the Mikulski Archive for Space Telescopes (MAST). This work has made use of data from the European Space Agency (ESA) mission {\it Gaia} (\url{https://www.cosmos.esa.int/gaia}), processed by the {\it Gaia} Data Processing and Analysis Consortium (DPAC, \url{https://www.cosmos.esa.int/web/gaia/dpac/consortium}). Funding for the DPAC has been provided by national institutions, in particular the institutions participating in the {\it Gaia} Multilateral Agreement. This publication makes use of data products from the Two Micron All Sky Survey, which is a joint project of the University of Massachusetts and the Infrared Processing and Analysis Center/California Institute of Technology, funded by the National Aeronautics and Space Administration and the National Science Foundation. Some of the observations in the paper made use of the High-Resolution Imaging instrument Zorro. Zorro was funded by the NASA Exoplanet Exploration Program and built at the NASA Ames Research Center by Steve B. Howell, Nic Scott, Elliott P. Horch, and Emmett Quigley. Zorro was mounted on the Gemini South telescope of the international Gemini Observatory, a program of NSF’s NOIRLab, which is managed by the Association of Universities for Research in Astronomy (AURA) under a cooperative agreement with the National Science Foundation on behalf of the Gemini partnership: the National Science Foundation (United States), National Research Council (Canada), Agencia Nacional de Investigaci\'on y Desarrollo (Chile), Ministerio de Ciencia, Tecnolog\'ia e Innovaci\'on (Argentina), Minist\'erio da Ci\^encia, Tecnologia, Inova\c{c}\~oes e Comunica\c{c}\~oes (Brazil), and Korea Astronomy and Space Science Institute (Republic of Korea). We thank the RECONS team (\url{http://www.recons.org/}) for CHIRON support. We thank the Cerro Tololo Inter-American Observatory (CTIO) TelOps team for MEarth support. This research made use of \textsf{exoplanet} \citep{exoplanet:joss,exoplanet:zenodo} and its dependencies \citep{exoplanet:agol20,exoplanet:arviz, exoplanet:astropy13, exoplanet:astropy18, exoplanet:luger18,exoplanet:pymc3, exoplanet:theano}.
\end{acknowledgments}

%% To help institutions obtain information on the effectiveness of their 
%% telescopes the AAS Journals has created a group of keywords for telescope 
%% facilities.
%
%% Following the acknowledgments section, use the following syntax and the
%% \facility{} or \facilities{} macros to list the keywords of facilities used 
%% in the research for the paper.  Each keyword is check against the master 
%% list during copy editing.  Individual instruments can be provided in 
%% parentheses, after the keyword, but they are not verified.

\vspace{5mm}
\facilities{MEarth, TESS, CTIO:1.5m (CHIRON), Gemini:South (Zorro)}

%% Similar to \facility{}, there is the optional \software command to allow 
%% authors a place to specify which programs were used during the creation of 
%% the manuscript. Authors should list each code and include either a
%% citation or url to the code inside ()s when available.

%%\software{astropy \citep{2013A&A...558A..33A,2018AJ....156..123A}
%%          }

\bibliography{main}{}
\bibliographystyle{aasjournal}

%% This command is needed to show the entire author+affiliation list when
%% the collaboration and author truncation commands are used.  It has to
%% go at the end of the manuscript.
%\allauthors

%% Include this line if you are using the \added, \replaced, \deleted
%% commands to see a summary list of all changes at the end of the article.
%\listofchanges

\end{document}